\documentclass [aps, prl, amssymb, twocolumn, showpacs]{revtex4-2}
\usepackage{graphicx,epsfig, amsfonts, amssymb, amsmath}
\usepackage{bm}
\usepackage{times}
\usepackage[colorlinks,citecolor=blue,linkcolor=red]{hyperref}
\usepackage{color}

\begin{document} 
 
\title{Geometric Thermoelectric Pump: Energy Harvesting beyond Seebeck and Pyroelectric Effects}

\author {Jie Ren}
\email{Corresponding address: xonics@tongji.edu.cn}
\affiliation{Center for Phononics and Thermal Energy Science, China-EU Joint Lab on Nanophononics, Shanghai Key Laboratory of Special Artificial Microstructure Materials and Technology,
School of Physics Science and Engineering, Tongji University, Shanghai 200092, China} 

\date{\today}

\begin{abstract}
Thermal-electric conversion is crucial for smart energy control and harvesting, such as thermal sensing and waste heat recovering. So far, people are aware of two main ways of direct thermal-electric conversion, Seebeck and pyroelectric effects, each with different working mechanisms, conditions and limitations.
Here, we report the concept of ``Geometric Thermoelectric Pump'', as the third way of thermal-electric conversion beyond Seebeck and pyroelectric effects.
In contrast to Seebeck effect that requires spatial temperature difference, Geometric Thermoelectric Pump converts the time-dependent ambient temperature fluctuation into electricity. Moreover, Geometric Thermoelectric Pump does not require polar materials but applies to general conducting systems, thus is also distinct from pyroelectric effect. We demonstrate that Geometric Thermoelectric Pump results from the temperature-fluctuation-induced charge redistribution, which has a deep connection to the topological geometric phase in non-Hermitian dynamics, as a consequence of the fundamental nonequilibrium thermodynamic geometry.
The findings advance our understanding of geometric phase induced multiple-physics-coupled pump effect and provide new means of thermal-electric energy harvesting. 
\end{abstract}






\maketitle

About 90 percent of the world's energy is utilized through heating and cooling, which makes
energy waste a great bottleneck to the sustainability of any modern economy~\cite{Chu}.
In addition to developing new technology of smart heat control~\cite{Li}, the global energy crisis can be alleviated by
recovering the wasted thermal energy.
In view of the inconvenient truth that more than 60\% of the energy utilization was lost mostly as wasted heat~\cite{60percent}, 
harvesting the thermal energy becomes critical to provide a cleaner and sustainable future~\cite{Chu}.

\begin{figure}
\hspace{-2mm}
\scalebox{0.35}[0.35]{\includegraphics{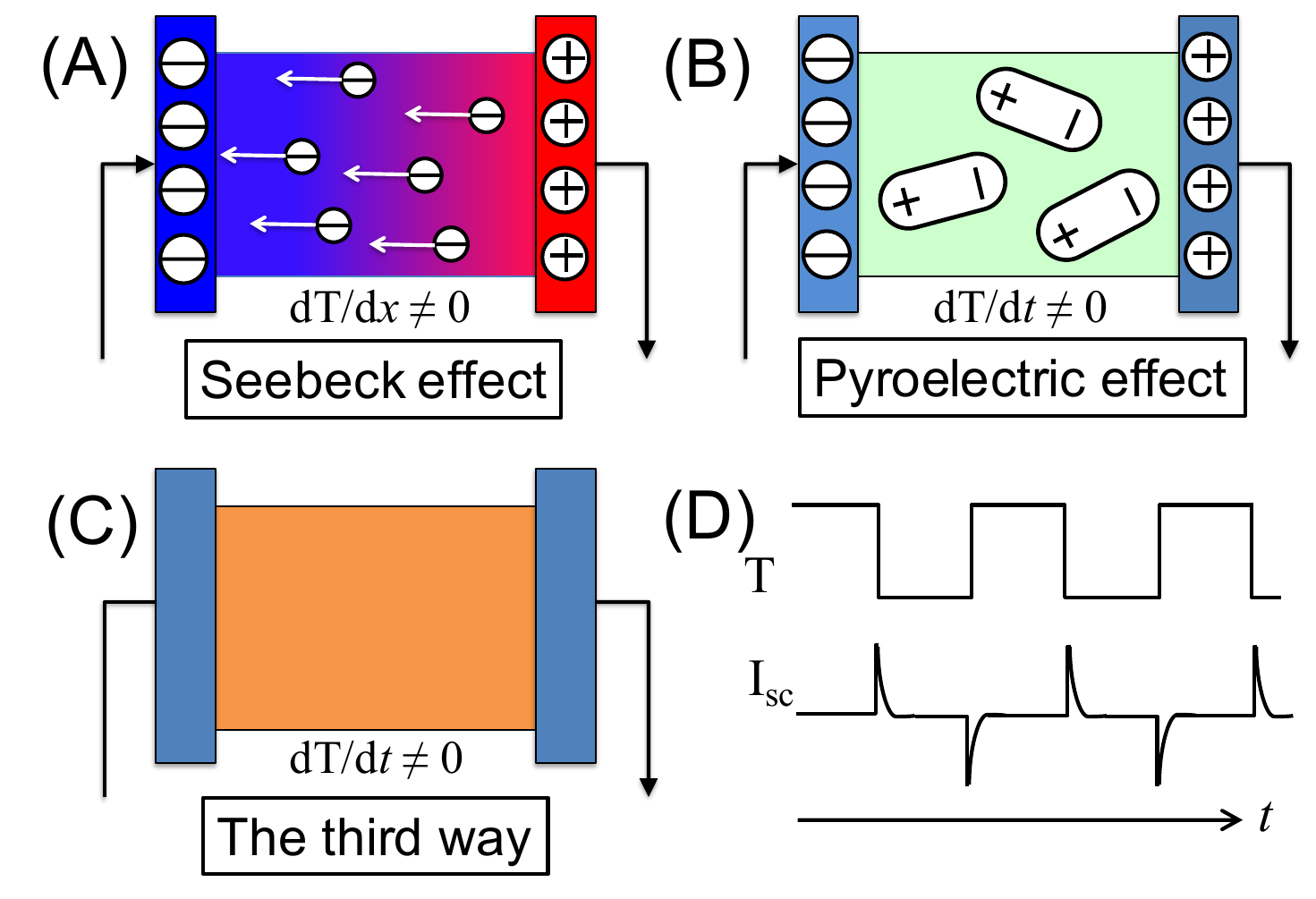}}
\vspace{-6mm}   
\caption{{\bf Comparison of different thermal-electric conversion ways.} 
({\bf A}) Seebeck effect converts the spatial temperature difference ($dT/dx\neq0$) to electricity by driving the thermal diffusion of charge carriers from hot to cold. 
({\bf B}) Pyroelectric effect converts the temporal temperature variation ($dT/dt\neq0$) to temporary electric current because variations of the temperature-dependent polarization in polar materials redistribute the surface charges. The typical behavior is similar to ({\bf D}), see also Refs.~\cite{ZLWang1,ZLWang2,ZLWang3}.
({\bf C}) Distinctly, the third way, i.e., geometric thermoelectric pump (GTEP), produces electric current from the time-dependent temperature fluctuation in contrast to Seebeck effect, and it does not require polar materials but applies to general conducting system in contrast to pyroelectric effect. 
({\bf D}) The typical behavior of temperature-fluctuation-induced temporary electric current by the GTEP. It is in a similar manner as that of pyroelectric effect, see also Refs.~\cite{ZLWang1,ZLWang2,ZLWang3}.
} 
\label{fig1}
\end{figure}

Thermal-electric energy harvesting mainly relies on two principles: Seebeck effect (Fig.~\ref{fig1}A) and pyroelectric effect (Fig.~\ref{fig1}B). 
The Seebeck effect utilizes the spatial temperature difference between two sides of materials to drive the diffusion of charge carriers so as to convert heat into electricity~\cite{DiSalvo,1note}. Besides recovering waste heat, Seebeck effect with its reciprocal has wide applications of cooling, heating, power generating~\cite{Bell} and thus has revitalized an upsurge of research interest recently~\cite{Tritt,Shakouri}.
However, when the ambient temperature is spatially uniform, we have to resort to the pyroelectric effect~\cite{Lang_book},  which utilizes the time-dependent temperature variation to convert heat into electricity but is restricted to pyroelectric materials~\cite{Whatmore}. This is due to the fact that the temporal temperature fluctuation modifies the spontaneous polarization of polar crystals, which consequently redistributes surface charges and produces temporary electric current~\cite{Lang_PT,2note}. In addition to thermal-electric energy harvesting~\cite{Sebald,Morozovska,ZLWang1,ZLWang2,ZLWang3}, pyroelectric effect has widespread applications in long-wavelength infrared sensing, motion detector, thermal image~\cite{Lang_PT}, and even nanoscale printing~\cite{Rogers}.

However, since Seebeck and pyroelectric effects have been known for quite a long time, one cannot help wondering: Does Nature only offer us these two main means for thermal-electric energy harvesting? Does any new principle of thermal-electric conversion exist beyond them? The Seebeck effect has been known for 200 years~\cite{Seebeck}, and the pyroelectric effect, named in 1824~\cite{Brewster}, can be even traced back to 314BC~\cite{Lang_PT}. Now, it is time to think ``out of the box'', as advocated by Arun Majumdar~\cite{Majumdar2}.

In this work, we describe a third way of thermal-electric energy harvesting beyond Seebeck effect and pyroelectric effect, coined as "Geometric Thermoelectric Pump". This third way converts the time-dependent ambient temperature fluctuation into electricity (Fig.~\ref{fig1}, C and D), thus in contrast to Seebeck effect that requires spatial temperature difference. The third way is also distinct from pyroelectric effect in the sense that it does not require polar materials but applies to general conducting systems, although they produce the electricity in a similar manner (Fig.~\ref{fig1}D). This third way of thermal-electric conversion (Geometric ThermoElectric Pump, GTEP for short) results from the charge redistribution through the temperature fluctuation and has a deep connection to the topological geometric phase in non-Hermitian dynamics. 

\begin{figure}
\hspace{-2mm}
\scalebox{0.45}[0.45]{\includegraphics{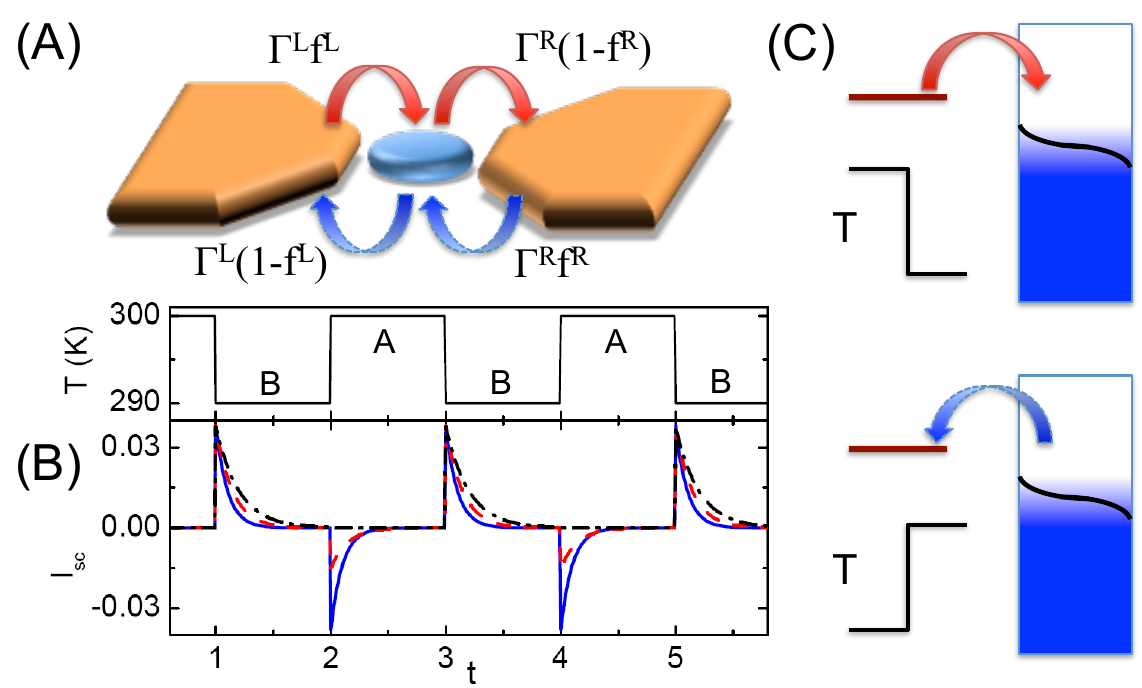}}
\vspace{-3mm}   
\caption{{\bf The principle of GTEP effecct of thermal-electric conversion from the ambient temperature fluctuation.} 
({\bf A}) Sketch of the single-level nanodevice and its charge transfer dynamics. 
({\bf B}) Temperature-fluctuation-induced temporary electric current. Temperature varies between $T_A=300$ K and $T_B=290$ K. Other parameters are $\varepsilon_0=40$ meV, $\Gamma^L_A=\Gamma^R_B=5$, $\Gamma^R_A=\Gamma^L_B=5$ (solid), $2$ (dash), $0$ (dash dot). The latter two cases show that the asymmetric system-lead coupling variation rectifies the ac current. 
({\bf C}) Schematic illustration of the electron transfer through the right lead under temperature variation: when temperature decreases, the electron transfers from the central level to the lead; while when temperature increases, the electron is injected from the lead to the central level. Results can be reversed if the system changes from hole-type ($\varepsilon_0>0$) to electron-type ($\varepsilon_0<0$). Similar phenomena can be found in bulk materials when transport changes from conduction band to valance band. GTEP effect will vanish in systems with electron-hole symmetry.
} 
\label{fig2}
\end{figure}

The GTEP effect is similar to the geometric heat pump~\cite{GHP1, GHP2} and is a consequence of the fundamental nonequilibrium thermodynamic geometry. 
To illustrate the GTEP effect, we consider a typical nanodevice with only one electron level, as shown in Fig.~\ref{fig2}A. Although we choose the single-level dot for demonstration due to the simplicity, it is worth emphasizing that the following discussions can be readily generalized to bulk materials. The mechanism of GTEP can be very general in periodically-driven (and even stochastically-driven) nonequilibrium systems with multiple-physics-coupled transports. 

The single-level nanodevice also has the great advantage of scalability and tunability, and it has many realizations such as single molecular junction, quantum dot, and quantum-point-contact system that have shown promise for energy applications~\cite{QD_TE,nano2,SMJ1,SMJ2}. The transfer kinetics is described as follows~\cite{ME1,ME2}: An electron can hop from the $v$ $(=L,R)$ lead into the single level with rate $\Gamma^vf^v$; When the level is occupied by an electron, the electron can escape to the $v$ lead with rate $\Gamma^v(1-f^v)$. Here $f^v=[e^{(\varepsilon_0-\mu^v)/(k_BT)}+1]^{-1}$ denotes the Fermi-Dirac distribution at the $v$ lead with $\varepsilon_0$ being the energy of the single level. $\Gamma^v$ is proportional to the system-lead coupling and the density of state of the lead $v$ at $\varepsilon_0$, so that $\Gamma^v$ is generally temperature-dependent~\cite{TDOS1,TDOS2}.

Periodic temperature fluctuation is imposed on the whole system, following Refs.~\cite{ZLWang1,ZLWang2,ZLWang3}.  
Without loss of generality, we adopt the square-wave fluctuation that is convenient for theoretical analysis and can be decomposed as linear superpositions of periodic sine functions.  The square-wave approximation is also justified since the temperature relaxation timescale in metallic leads can be very fast, even up to picosecond. Thus, once the fluctuation period is much larger than the relaxation time in metallic leads, the fluctuation can be safely approximated as an abrupt  change. Moreover, we checked the fluctuation with smooth change, e.g., a single-frequent sine or cosine function. The results do not change qualitatively. In fact, any smooth (and even stochastic) temperature change protocols can be trotterized (decomposed) into step-wise discretized protocol. 

As such, the ambient temperature switches between high temperature $T_A$ and low temperature $T_B$, each with duration time $\mathcal T_p$, as depicted in Fig.~\ref{fig2}B. Accordingly, the system's evolution can be divided into two ensembles, of which the two sets of parameters are denoted by the corresponding superscript $A$ and $B$, respectively.  
Without doubt, in either of two ensembles the system is in equilibrium with a homogenous temperature, which as is well understood can not produce electric current. However, as we will 
see 
the electric current indeed emerges with the help of temperature fluctuations, because switching between two different equilibrium ensembles ultimately makes the whole evolution out of equilibrium. We set $\mu^v=0$ so that $f^L_{A,B}=f^R_{A,B}=f_{A,B}$ and focus on the short-circuit current through the right lead. 
Straightforward calculations lead us to the analytic expressions of temporary (short-circuit) electric currents:
\begin{eqnarray}
I_{sc}=\left\{
\begin{array}{l}
-\Gamma^R_A(f_A-f_B)\frac{1-e^{-K_B\mathcal T_p}}{1-e^{-(K_A+K_B)\mathcal T_p}}e^{-K_A \tau_A},  \\
 \Gamma^R_B(f_A-f_B)\frac{1-e^{-K_A\mathcal T_p}}{1-e^{-(K_A+K_B)\mathcal T_p}}e^{-K_B \tau_B},  
\end{array}
\right.
\label{eq:It}
\end{eqnarray}
where $K_u=\Gamma^L_u+\Gamma^R_u$ and $\tau_u=\mod(t\in u, \mathcal T_p)$ with $u=A,B$. The first (second) line of $I_{sc}$ is the evolution expression of temporary electric current when $t\in A (B)$ within the period $A (B)$.
Results of the GTEP effect are displayed in Fig.~\ref{fig2}B, which shows consistency with the thermal-electric generation of pyroelectric effect in a similar manner~\cite{ZLWang1,ZLWang2,ZLWang3}.

The kinetics 
of the GTEP induced thermal-electric conversion can be understood as follows (see also Fig.~\ref{fig2}C):
Before the temperature change, the central system and the lead have built an equilibrium, thus no net charge transfer. However, when ambient temperature fluctuates to lower one, less electrons will populate above the lead's Fermi level since the thermal broadening will shrink. As such,  the balance is broken and the electron tends to transfer from the central system to the lead and then relax to a new equilibrium.
When temperature fluctuates back to higher one, more electrons will populate above the lead's Fermi level due to the thermal broadening. Thus, the redundant conduction electron in the lead will transfer to the central system, trying to build a new balance. Therefore, even without polar materials, the GTEP effect is able to generate electric current from the ambient temperature fluctuation in general conducting system, which is intrinsic to the nonequilibrium kinetics. 
Although with different principles, considering the similar generated electric current behaviors between the GTEP and pyroelectric effect~\cite{ZLWang1,ZLWang2,ZLWang3}, we may use a full-wave bridge circuit to rectify the ac current~\cite{ZLWang2}, which also can be achieved by the asymmetric system-lead coupling variation (Fig.~\ref{fig2}B). 

So far, we have shown that GTEP effect, the-third-way of thermal-electric conversion, is a consequence of the  charge redistribution through the temporal temperature fluctuation. In below, we are going to unravel its deep connection to the topological geometric phase in non-Hermitian quantum mechanics, as a consequence of the nonequilibrium thermodynamic geometry. 
The evolution of the system can be described by a Schr\"odingier equation in imaginary time, that is, a twisted master equation with the counting field $\chi$~\cite{GHP1}:
\begin{equation}
\frac{d}{dt}|\Psi_{\chi}(t)\rangle= {\hat H}(\chi,t)|\Psi_{\chi}(t)\rangle,
\label{eq:Heff}
\end{equation}
where $|\Psi_{\chi}(t)\rangle=\sum^{\infty}_{q=-\infty}e^{iq\chi}[p_0(q,t), p_1(q,t)]^{\text T}$ with $p_{0/1}(q,t)$ denoting two joint probabilities that at time $t$ the single level is empty or occupied by an electron while there have already $q$ electrons transferred into the right lead. The transfer operation $\hat{H}$ is reminiscent of the Hamiltonian but non-Hermitian. In either $u=A$ or $B$ equilibrium ensemble, ${\hat H}$ is expressed as:
${\hat H}_u=
\left(
\begin{array}{cc}
-K_uf_u&  (\Gamma^L_u+\Gamma^R_ue^{i\chi})(1-f_u)\\
 (\Gamma^L_u+\Gamma^R_ue^{-i\chi})f_u& -K_u(1-f_u)
\end{array}
\right)$.
Therefore, after $n$ fluctuation periods the evolution at time $t=2n\mathcal T_p$ is described by the characteristic function:
\begin{equation}
Z_{\chi}=\langle\bm1|\left[e^{\hat{H}_B\mathcal T_p}e^{\hat{H}_A\mathcal T_p}\right]^n|\Psi_{\chi}(0)\rangle\sim e^{n\Phi(\chi)}
\label{eq:Z}
\end{equation}
with $\langle\bm1|=(1,1)$, 
which in turn generates the total electron transfer per period via the relation:
$Q:=\left.{\partial \Phi(\chi)}/{\partial (i\chi)}\right|_{\chi=0}$ with the cumulant generating function defined as $\Phi(\chi):=\lim_{n\rightarrow\infty}\frac{1}{n}\ln Z_{\chi}$.

At the slow switch limit (large $\mathcal T_p$), the evolution at either ensemble $\hat{H}_{u}$ ($u=A$ or $B$) is dominated by the corresponding eigenmode, of which the eigenvalue $\lambda_u$ has the largest real part, with $|\psi_u\rangle$ and $\langle\varphi_u|$ the corresponding normalized right and left eigenvectors, such that $e^{\hat{H}_u\mathcal T_p}\approx|\psi_u\rangle e^{\lambda_u\mathcal T_p}\langle\varphi_u|$. Therefore, the characteristic function can be expanded as:
\begin{equation}
Z_{\chi}(t)\approx e^{(\lambda_A+\lambda_B)n\mathcal T_p}[\langle\varphi_A|\psi_B\rangle\langle\varphi_B|\psi_A\rangle]^nC.
\end{equation}
with $C={\langle\bm1|\psi_B\rangle\langle\varphi_A|\Psi_{\chi}(0)\rangle}/{\langle\varphi_A|\psi_B\rangle}$ an unimportant coefficient.
This expansion of $Z_{\chi}$ gives the cumulant generating function $\Phi(\chi)$, composed of two contributions: $Z_{\chi}\sim e^{n\Phi(\chi)}=e^{n(\Phi_{\text{dyn}}+\Phi_{\text{top}})}$, where 
\begin{subequations}
\begin{align}
\Phi_{\text{dyn}}&=(\lambda_A+\lambda_B)\mathcal T_p,   \\
\Phi_{\text{top}}&= \ln\left[\langle\varphi_A|\psi_B\rangle\langle\varphi_B|\psi_A\rangle\right].
\label{eq:phitopo}
\end{align}
\end{subequations}
Being reminiscent of the dynamic phase in quantum mechanics, we call $\Phi_{\text{dyn}}(\chi)$ the dynamic phase contribution of the cumulant generating function, which presents the average evolution as if the system evolves in either ensemble separately and then do a simple summation. As a consequence of the detailed balance at each equilibrium ensemble, the dynamic phase contribution to the electric-thermal conversion is always zero: $\partial\Phi_{\text{dyn}}/\partial(i\chi)|_{\chi=0}=0$.

The second part $\Phi_{\text{top}}(\chi)$ however possesses a nontrivial topological geometric phase interpretation, which is responsible for the temperature-fluctuation-induced electron transfer. Different from the Hermitian Hamiltonian in quantum mechanics, the twisted transfer operator $\hat{H}_u$ are non-Hermitian such that its left eigenvector $\langle\varphi_u|$ is not the Hermitian conjugate of the corresponding right eigenvector $|\psi_u\rangle$, but only bi-orhonormal with each other.
Therefore, writing $\langle\varphi_A|\psi_B\rangle=|\langle\varphi_A|\psi_B\rangle|e^{i\phi_{AB}(\chi)}$ and $\langle\varphi_B|\psi_A\rangle=|\langle\varphi_B|\psi_A\rangle|e^{i\phi_{BA}(\chi)}$ with angle $\phi_{AB}$ and $\phi_{BA}$ as the analogy of Pancharatnam's topological phase between discrete states~\cite{Pan_phase1,Pan_phase2}, we have $\phi_{AB}\neq\phi_{BA}$ generally.
As such, the transferred electron number per period resulting from this topological phase contribution is obtained as:
\begin{eqnarray}
Q&=&\left.\frac{\partial\Phi_{\text{top}}}{\partial(i\chi)}\right|_{\chi=0}=
\frac{(f_A-f_B)(\Gamma^L_A\Gamma^R_B-\Gamma^L_B\Gamma^R_A)}{K_AK_B}.
\label{eq:topophase}
\end{eqnarray}
This quantity only depends on the topological phase properties of $\langle\varphi_u|$ and $|\psi_u\rangle$, the left and right ground states of the non-Hermitian dynamics $\hat H_u$, and is independent on the period in the large $\mathcal T_p$ limit, which looks similar to the behavior governed by the adiabatic Berry phase effect~\cite{Sinitsyn1,Sinitsyn2,Ren1,Ren2,Ren3,Ren4,Berry1,Berry2} with smooth and continuous driving protocols. 

However, we note that here the Pancharatnam-like topological phase contribution is distinct from previous Berry phase studies~\cite{Ren1,Ren2,Ren3,Ren4,Berry3,Berry4,Berry5}, where the temperature modulations had a phase lag in order to form a closed loop with non-zero area in the parameter space for the latter case. The Pancharatnam-like phase here actually denotes an isothermal non-adiabatic process due to the abruptly discrete switching. 
The nonzero net charge transfer per period is kinetically offered by the nonequilibrium asymmetric relaxations between two equilibrium ensembles $A\leftrightarrow{B}$ with different temperatures, which is characterized by the non-cancellation of phases of inner products 
$\langle\varphi_B|\psi_A\rangle$ and $\langle\varphi_A|\psi_B\rangle$ in the topological phase contribution $\Phi_{\text{top}}$.
This is distinct from those in quantum mechanics, where if having only two-state switching, $\phi_{AB}$ will be always equal to $\phi_{BA}$~\cite{Pan_phase2}, leading to zero contribution. In Hermitian quantum mechanics, to avoid the cancellation of phases, at least three-state change was necessary to form a closed loop in the parameter space with nonzero area in order to get the nonzero contribution. 
Even in the continuous driven master equation with counting field, a nonzero area enclosed by the driving protocol in the parameter space was necessary to obtain a finite Berry's geometric phase contribution.
Nevertheless, two-state switch here is already enough to generate finite charge transfer from the Pancharatnam-like phase through Eqs.~(\ref{eq:phitopo}) and (\ref{eq:topophase}),  even though the closed area of the driven protocol is zero in the parameter space, because for non-Hermitian dynamics $\langle\varphi_u|$ is not the Hermitian conjugate of the corresponding right eigenvector $|\psi_u\rangle$. Similar phenomena has also been demonstrated in temporally-driven macroscopic thermal diffusion~\cite{GHP2} in both theory and experiment. In Ref.~\cite{GHP2},  the connection between Trotterized discrete switching protocol (Pancharatnam's phase induced geometric heat pump) and continuous driving protocol (Berry's phase induced geometric heat pump) are discussed in details.  

\begin{figure}
\vspace{-3mm} 
\hspace{-5mm}
\scalebox{0.55}[0.55]{\includegraphics{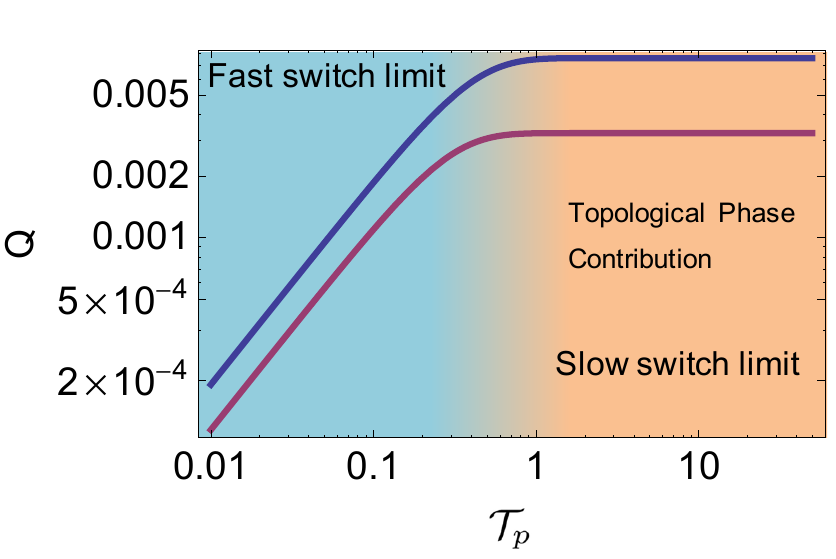}}
\vspace{-5mm}   
\caption{{\bf The electron transfer generated per period by the GTEP effect of thermal-electric energy harvesting.} 
The thermal-electric conversion in the slow switch limit has an interpretation of topological phase, and gives the upper bound of the thermal-electric conversion per period. $\Gamma^R_A=\Gamma^L_B=0$ (upper curve), $2$ (lower curve). Other parameters are the same as used in Fig.~\ref{fig2}. For the maximal asymmetric coupling variation, the topological geometric phase contribution will reach a plateau $Q=f_A-f_B$ at large $\mathcal T_p$, so that at large temperature fluctuation $f_A\approx1/2, f_B\approx0$, the geometric thermo-electric pumped charge will be half-quantized as $Q=1/2$.
} 
\label{fig3}
\end{figure}

At the fast switch limit (small $\mathcal T_p$), we can approximate $\exp{[\hat{H}_B\mathcal T_p]}\exp{[\hat{H}_A\mathcal T_p]}\approx \exp{[(\hat{H}_A+\hat{H}_B)\mathcal T_p]}$. Denote $\lambda_{AB}(\chi)$ as the dominated eigenvalue of the matrix $(\hat{H}_A+\hat{H}_B)$, the solution at fast switch limit is obtained as: 
 \begin{equation}
Q=\mathcal T_p\frac{\partial \lambda_{AB}}{\partial(i\chi)}=\frac{\mathcal T_p(f_A-f_B)(\Gamma^L_A\Gamma^R_B-\Gamma^L_B\Gamma^R_A)}{K_A+K_B}.
\label{eq:fastQ}
 \end{equation} 
Clearly, in this fast switch regime, $Q$ is proportional to the half-period $\mathcal T_p$, while the average current $Q/(2\mathcal T_p)$becomes a period-independent constant. 
 
In fact, by considering the characteristic function and the cumulant generating function in Eq. (\ref{eq:Z}), the full exact solution of the total thermally generated electron transfer per period at arbitrary switch rate (i.e., arbitrary $\mathcal T_p$) can be obtained as
\begin{eqnarray}
\label{eq:fullQ}
&&Q= \frac{\partial \Phi(\chi)}{\partial (i\chi)}|_{\chi=0}=\lim_{n\rightarrow\infty}\frac{1}{n}\frac{\partial\ln Z_{\chi}}{\partial (i\chi)}|_{\chi=0} \\
&=&\frac{(f_A-f_B)(\Gamma^L_A\Gamma^R_B-\Gamma^L_B\Gamma^R_A)}{K_AK_B}\frac{(1-e^{-K_A\mathcal T_p})(1-e^{-K_B\mathcal T_p})}{(1-e^{-(K_A+K_B)\mathcal T_p})}, \nonumber
\end{eqnarray}
which is also seen as the result from integrating Eq.~(\ref{eq:It}) within one period: $Q=\int_{0}^{2\mathcal T_p}dt I_{sc}(t)$. Clearly, the relaxation factor $\frac{(1-e^{-K_A\mathcal T_p})(1-e^{-K_B\mathcal T_p})}{(1-e^{-(K_A+K_B)\mathcal T_p})}$ plays the key role in the transition behavior from slow-switch limit ($\approx 1$ when $\mathcal T_p\gg 1/K_A, 1/K_B$) to fast-switch limit ($\approx \frac{K_AK_B}{K_A+K_B}\mathcal T_p$, when $\mathcal T_p\ll 1/(K_A+K_B)$).  
Therefore, the full exact solution in the small $\mathcal T_p$ limit coincides with the fast switch result and in the large $\mathcal T_p$ limit it reduces to the Pancharatnam-like topological-phase contribution Eq.~(\ref{eq:topophase}). 
Therefore, the full exact solution in the large $\mathcal T_p$ limit reduces to the Pancharatnam-like topological-phase contribution Eq.~(\ref{eq:topophase}) and  in the large $\mathcal T_p$ limit it coincides with the fast switch result Eq.~(\ref{eq:fastQ}).

As shown in Fig.~\ref{fig3},  the topological-phase contribution in the slow switching gives the upper limit of the thermal-electric conversion per period by the GTEP effect. Intuitively, $Q$ is proportional to $f_A-f_B$ that is in turn proportional to $T_A-T_B$, i.e., large temperature fluctuations produce more electricity. Moreover, Eq.~(\ref{eq:fullQ}) shows $Q$ is proportional to $\Gamma^L_A\Gamma^R_B-\Gamma^L_B\Gamma^R_A$. This indicates that the larger asymmetric coupling variation $\Gamma^L_A/\Gamma^R_A\neq\Gamma^L_B/\Gamma^R_B$, the larger the electric generation. The largest asymmetry could be achieved by letting $\Gamma^R_A=0$ in the first half period, and $\Gamma^L_B=0$ in the second half period, see Fig.~\ref{fig3}, which corresponds to the anti-phase transfer: in ensemble $A$ the electron transfers from the left lead to the central system that decouples with the right lead (with probability $f_A$); in ensemble $B$, the central system decouples with the left lead and the electron transfers from the central part to the right lead (with probability $f_B$). As such, the geometric thermo-electric pumped charge is $Q=f_A-f_B$ at slow switch limit, which will be half-quantized as $Q=1/2$ when temperature fluctuation is large ($f_A\approx1/2$ at high temperature and $f_B=0$ at low temperature). This picture also coincides with the largest rectification shown in Fig.~\ref{fig2}B and is sketched in Fig.~\ref{fig2}C. Qualitatively similar effect has also been demonstrated in temporally-driven macroscopic thermal diffusion both theoretically and experimentally~\cite{GHP2}. 

The same analysis can be applied to the nonequilibrium entropy production. Following Refs.~\cite{entropy1,entropy2,Seifert}, we can split the total entropy production into the system entropy production and the flux entropy production (i.e., the environment entropy production due to the dissipation of the transition flux), as $\dot S_{\text{tot}}=\dot S+\dot S_{\text{flux}}\geq0$, where the second equality holds only in equilibrium in the absence of net transferred electric flux. After one full switching period $2\mathcal T_p$ we see that the entropy change $\Delta{S}=\int^{2\mathcal T_p}_0 dt \dot S=0$ because the system's Gibbs entropy $S=-k_B\sum_{i}p_i\ln{p_i}$ does not change after a whole period so that $\Delta{S}_{\text{tot}}=\Delta{S}_{\text{flux}}=\int^{2\mathcal T_p}_0 dt \dot S_{\text{flux}}$.  
Therefore, to get the fluctuation information of the total entropy, we can then apply the full counting statistics to count the flux entropy $s_{\text{flux}}=\ln{\frac{\Gamma^v_u(1-f^v_u)}{\Gamma^v_uf^v_u}}=\varepsilon_0/T_u$ produced on each transition from the center level to both leads $v=L, R$ in each ensemble $u=A, B$. 

\begin{figure}
\hspace{-2mm}
\scalebox{0.75}[0.75]{\includegraphics{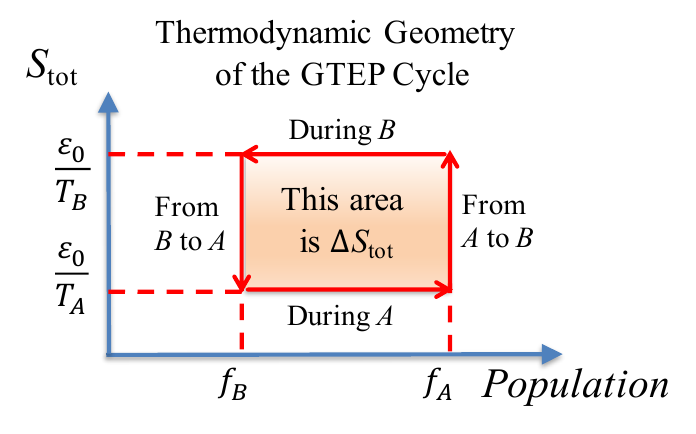}}
\vspace{-8mm}   
\caption{{\bf Thermodynamic geometry of entropy production in the GTEP cycle in slow switch limit.} 
The switch protocol forms a closed loop: when switching from ensemble $B$ to $A$, the system's entropy switches from $\varepsilon_0/T_B$ to $\varepsilon_0/T_A$, while the population of the central level starts to relax from $f_B$ to $f_A$ during stage $A$ in the large period limit; then when the system switches from $A$ to $B$, the system's entropy changes from $\varepsilon_0/T_A$ to $\varepsilon_0/T_B$ and the population of the central level will relax from $f_A$ to $f_B$; so that the evolution of GTEP forms a closed thermodynamic cycle, whose area is $\Delta{S}_{\text{tot}}=\varepsilon_0(\frac{1}{T_A}-\frac{1}{T_B})(f_A-f_B)$, exactly the same as the entropy production from the full counting statistics. 
} 
\label{fig4}
\end{figure}

In this way,  
the twisted transfer matrix with entropy counting becomes ${\hat H}_u=
\left(
\begin{array}{cc}
-K_uf_u&  (\Gamma^L_u+\Gamma^R_u)(1-f_u)e^{i\chi s_{\text{flux}}}\\
 (\Gamma^L_u+\Gamma^R_u)f_ue^{-i\chi s_{\text{flux}}} & -K_u(1-f_u)
\end{array}
\right)$. 
Similar to approaches from Eq.~(\ref{eq:Heff}) to Eq.~(\ref{eq:fullQ}) we obtain the total entropy change per period $\Delta{S}_{\text{tot}}=\Delta S_{\text{flux}}=\langle{s}_{\text{flux}}(2\mathcal T_p)\rangle-\langle{s}_{\text{flux}}(0)\rangle$, as: 
\begin{equation}
\Delta S_{\text{tot}}=\varepsilon_0\frac{(f_A-f_B)(T_A-T_B)}{T_AT_B}\frac{(1-e^{-K_A\mathcal T_p})(1-e^{-K_B\mathcal T_p})}{(1-e^{-(K_A+K_B)\mathcal T_p})}.
\label{eq:entropy}
\end{equation}
This total entropy change shares the same behavior of transferred electron number Eq.~(\ref{eq:fullQ}), as depicted in Fig.~\ref{fig3}, so we do not repeat the plotting here.  
Similarly, at slow switch limit $\mathcal T_p\rightarrow \infty$, the entropy change per period is
\begin{equation}
\Delta{S}_{\text{tot}}=\frac{\varepsilon_0(f_A-f_B)(T_A-T_B)}{T_AT_B},
\label{eq:topoS}
\end{equation}
which is independent of period and has the same topological interpretation of Pancharatnam's geometric phase as that of the thermoelectric pumped electron. The nonequilibrium thermodynamic geometry is clearly illustrated in Fig.~\ref{fig4} where the entropy production is exactly the area enclosed by the thermodynamic cycle of GTEP. At fast switch limit $\mathcal T_p\rightarrow 0$, the entropy change per period is
\begin{equation}
\Delta S_{\text{tot}}=\frac{\varepsilon_0K_AK_B(f_A-f_B)(T_A-T_B)}{(K_A+K_B)T_AT_B}\mathcal T_p,
\label{eq:fastS}
\end{equation}
proportional to the period, so that the average entropy production rate $\Delta S_{\text{tot}}/(2\mathcal T_p)$ is a period-independent constant. These similar behaviors and underlying physics between  Eq.~(\ref{eq:fullQ}) and Eq.~(\ref{eq:entropy}) (Eq.~(\ref{eq:topophase}) and Eq.~(\ref{eq:topoS}), Eq.~(\ref{eq:fastQ}) and Eq.~(\ref{eq:fastS}) ) indicate that the more entropy dissipated into the environment (metallic leads), the farther the system will be pulled away from the equilibrium (more entropy changes), and thus the more electron transfers can be generated.

We now briefly estimate the open-circuit voltage, $V_{oc}$. Clearly, the maximum $V_{oc}$ will be achieved in the largest asymmetric case $\Gamma^L_A/\Gamma^R_A\gg\Gamma^L_B/\Gamma^R_B$. Let's further assume $+V_{oc}/2$ is imposed on the left lead and $-V_{oc}/2$ is imposed on the right one, and recall $T_A>T_B$. In the first half period, ensemble $A$, the electron transfers from the left lead to the central system ($\Gamma^R_A=0$), carrying entropy $(\varepsilon_0+eV_{oc}/2)/T_A$; in the second half period, ensemble $B$, the electron transfers from the central part to the right lead ($\Gamma^L_B=0$), releasing entropy $(\varepsilon_0-eV_{oc}/2)/T_B$. The reversibility of the charge transfer at the open-circuit condition requires zero entropy production, so that $(\varepsilon_0+eV_{oc}/2)/T_A=(\varepsilon_0-eV_{oc}/2)/T_B$. Therefore, we have the relation
\begin{equation}
V_{oc}=2\frac{\varepsilon_0}{e}\frac{T_A-T_B}{T_A+T_B},
\label{eq:Voc}
\end{equation}
which is also confirmed numerically (not shown here). It indicates that increasing the temperature fluctuation and meanwhile decreasing the average temperature can increase the open-circuit voltage, which is upper bounded by $V^{\text{max}}_{oc}=2\varepsilon_0/e$.

Finally, we provide realisitic parameter estimates for the GTEP effect converted electricity. The real coupling rates are usually at the scale of GHz, indicating one electron hopping per nanosecond. We then can confer the unit GHz to the dimensionless couplings used for calculating Figs.~\ref{fig2} and \ref{fig3}. As such the peak current $I_{sc}=0.04$ in Fig.~\ref{fig2} corresponds to $0.04*1.6\times10^{-19}$C$/10^{-9}$s$=6.4$ pA, and the average current per period at $Q/(2\mathcal T_p)\approx0.004$ in Fig.~\ref{fig3} corresponds to $0.64$ pA. Considering the advantage of scalability in nanodevices, if we pack molecular quantum dots with spacing $50$ nm between two metallic plate leads, we will have the packing density $4\times10^{10}$/cm$^2$ that leads to the significant current density around the order of $0.1$A/cm$^2$. Also, the output voltage can be further optimized in practice, e.g. in the molecular junction the energy gap between the LUMO and the lead's Fermi level is around $\varepsilon_0=2$ eV~\cite{SMJ1,SMJ2}, which compared to our used $\varepsilon_0=40$ meV can further improve the output voltage according to Eq.~(\ref{eq:Voc}). Moreover, the temperature fluctuation in the industry or deep space, e.g., the satellite or the space station, can be as large as from 100 K to 500 K, which can generate even better electricity by the GTEP effect.
 
Although there is widespread demand in thermal energy science for both cost reductions and performance improvements, it is important at this stage to explore ``out of the box" in principle. The GTEP effect reported here as the-third-way of thermal-electric conversion is such an attempt. In contrast to Seebeck and pyroelectric effects, GTEP effect is a consequence of the temperature-fluctuation-induced charge redistribution. This-third-way of thermal-electric conversion does not require neither polarized materials nor spatial temperature heterogeneity. It results from the fundamental nonequilibrium thermodynamic geometry~\cite{GHP1, GHP2} and has a deep connection to the topological geometric phase in non-Hermitian quantum mechanics.  

Future extensions include the smooth and continuous temperature variation, macroscopic version of geometric thermoelectric pump effect in bulk materials, thermo-spin conversion and pumping~\cite{SSE1,SSE2,SSE3,SSE4}, and complex-network-based thermal energy conversion~\cite{cyclenetwork}, where more detailed power and efficiency analysis should be desired. 
Also, exploring the geometric-phase-induced pumping effect in spatiotemporal metamaterials will be of great potential interest~\cite{JP1,JP2,JP3}.
Moreover, multi-physics coupled transport will provide versatile opportunities for new types of thermo-electric conversion and vice versa.  
For example,  Ref.~\cite{YuanYang} requires an electrochemical system combined with temperature fluctuations for heat energy harvesting. And based on the multi-caloric effect, such as elastocaloric effect~\cite{ELC}, electrocaloric effect~\cite{EC1,EC2,EC3,EC4,EC5,EC6,EC7}, many novel heat pumping devices based on periodic modulation have been proposed. 
The present work, however, depends neither on electrochemical systems nor on pyroelectric or other specific multi-caloric materials, but lays the foundation of understanding such temperature-fluctuation-induced electricity in nonequilibrium systems in a general way.  
We believe that, as human's ability to design and manipulate nano-systems is improving, the ``Geometric ThermoElectric Pump'' (GTEP) effect, as the-third-way of thermal-electric conversion, will provide the new means of energy harvesting to harness the ubiquitous temperature fluctuations in the world.

\begin{acknowledgments}
Jie Ren acknowledges the support from the National Natural Science Foundation of China (Grant Nos. 11935010), the Natural Science Foundation of Shanghai (Grant Nos. 23ZR1481200, 23XD1423800), and the Opening Project of Shanghai Key Laboratory of Special Artificial Microstructure Materials and Technology.
\end{acknowledgments}

\bibliography{ref.bib}

\begin{thebibliography}{64}
\bibitem [{1no()}]{1note}%
  \BibitemOpen
  \href@noop {} {\bibinfo  {journal} {Spin Seebeck effect is also included
  here, which first converts the thermal difference into spin current and then
  into electric current, through the inverse spin Hall effect}\ }\BibitemShut
  {NoStop}%
\end{thebibliography}

\end{document}